\documentstyle[12pt,epsfig,fleqn,amssymb,citesort]{article}

\newcommand{\be}{\begin{eqnarray}}
\newcommand{\ee}{\end{eqnarray}}
\newcommand{\nee}{\nonumber\end{eqnarray}}
\newcommand{\nn}{\nonumber}

\newcommand{\eq}[1]  {\mbox{(\ref{eq:#1})}}

\def\figa
{
\begin{figure}[t]\vspace{-0.5cm}
\hspace{-1cm}
\epsfig{figure=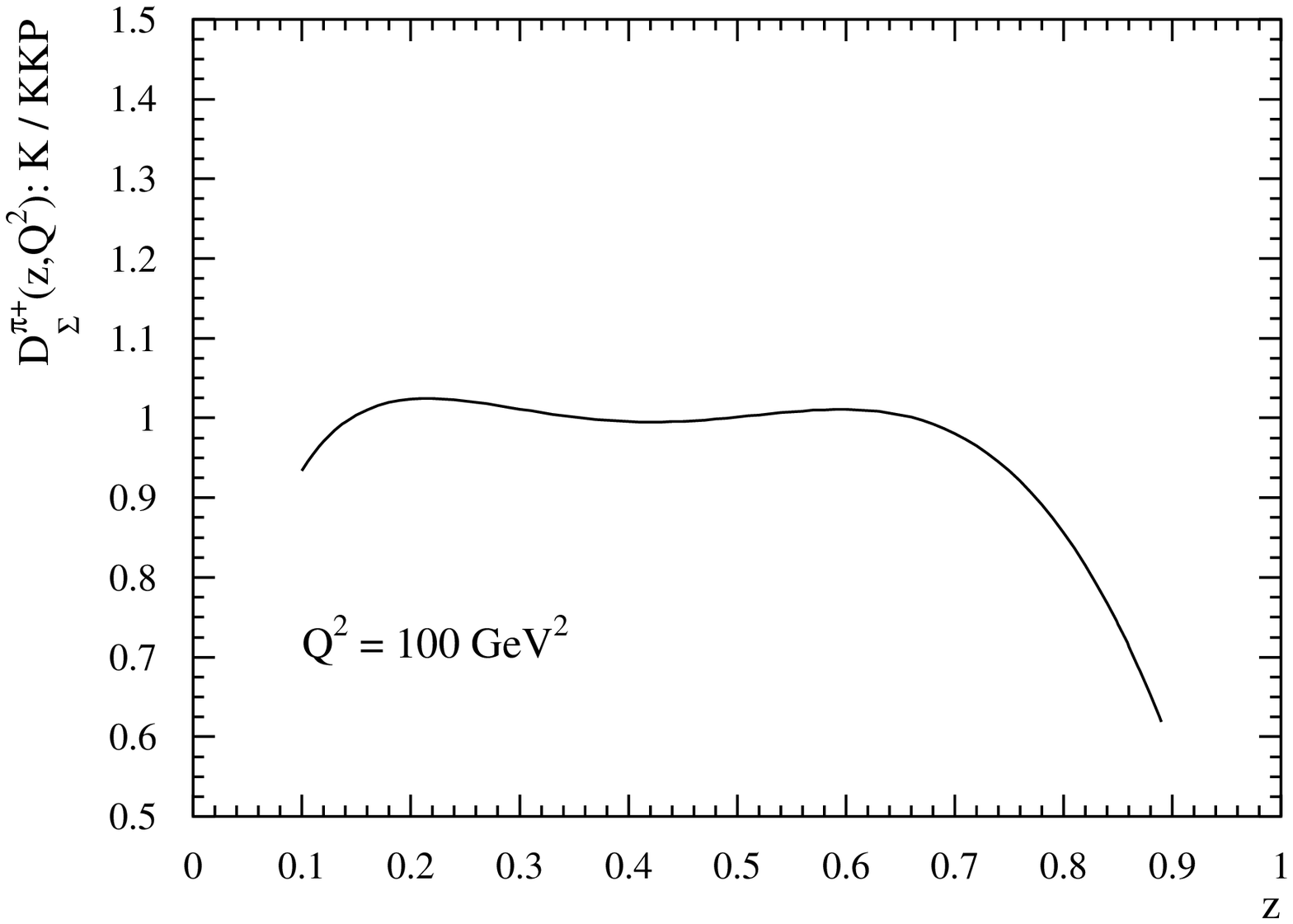,width=8.5cm}
\epsfig{figure=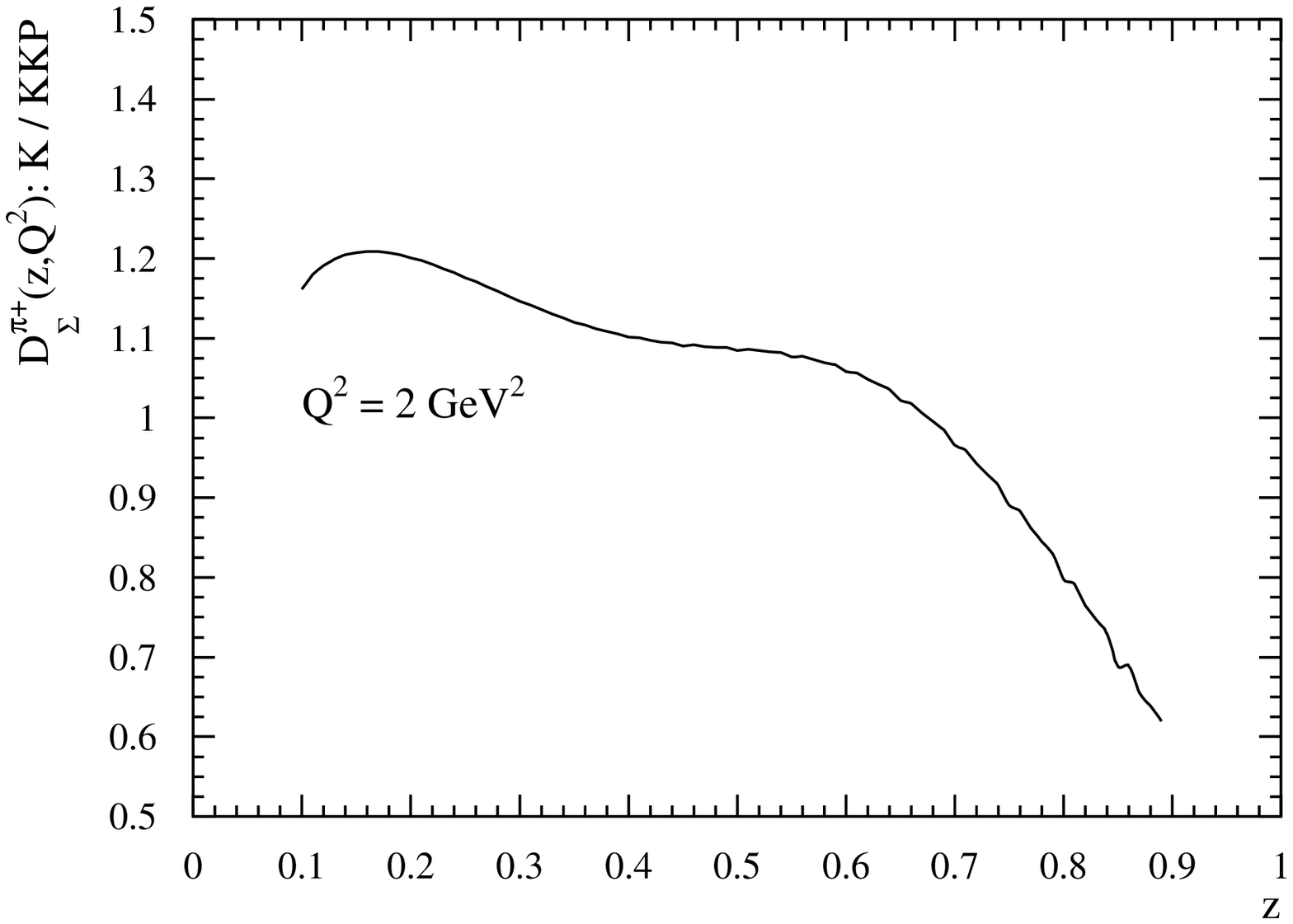,width=8.5cm}
\vspace{-0.5cm}
\caption{
The ratio of singlet fragmentation functions $D_{\Sigma}^{\pi^+}$
found in \cite{Kr} (K) and \cite{KKP} (KKP), 
at  $Q^2 = 100\ {\rm GeV}^2$  (left) and at a typical SIDIS
value $Q^2 = 2\ {\rm GeV}^2$  (right).
}
 \label{fig1}
\end{figure}
}

\def\figb
{
\begin{figure}[t]
\vspace{-1cm}
\hspace{2cm}
\epsfig{figure=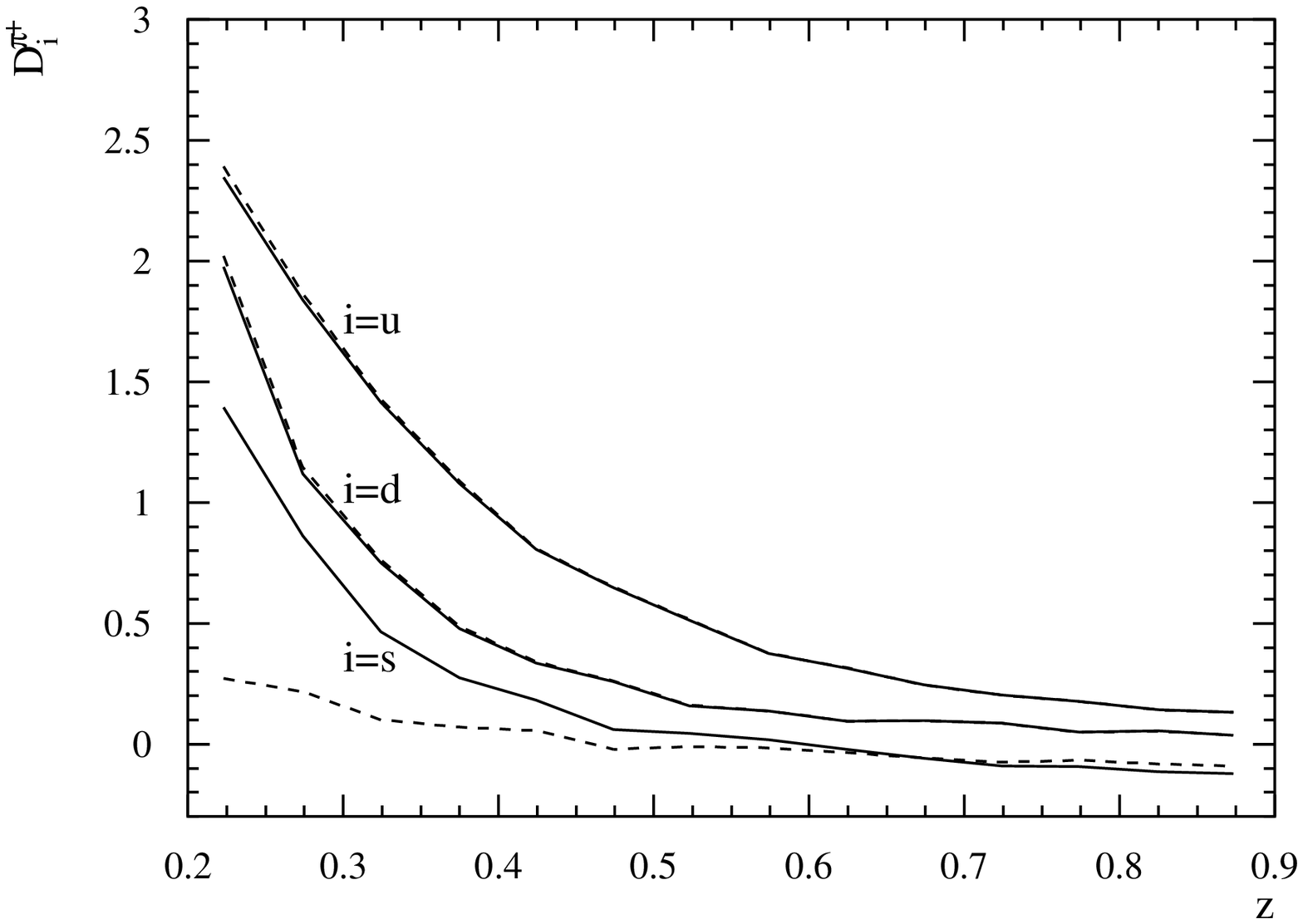,width=15cm}
\vspace{-1cm}
\caption{
The valence-type ($D_u^{\pi^+}$), sea-type ($D_d^{\pi^+}$) and
strange-type ($D_s^{\pi^+}$) FF into charged pions extracted
under the assumption of isospin-invariance  from HERMES SIDIS
measurements supplemented by the singlet FF of \cite{Kr} (solid)
and \cite{KKP} (dashed), respectively. Details are given in the
text. Switching between the LO $\leftrightarrow$ NLO
parametrizations of \cite{Kr,KKP} leads to similar variations in
the FFs whereas the variation from employing different
unpolarized PDFs is negligible.
}
\label{fig2}
\end{figure}
}

\def\figc
{
\begin{figure}[t]
\vspace{-1cm}
\hspace{2cm}
\epsfig{figure=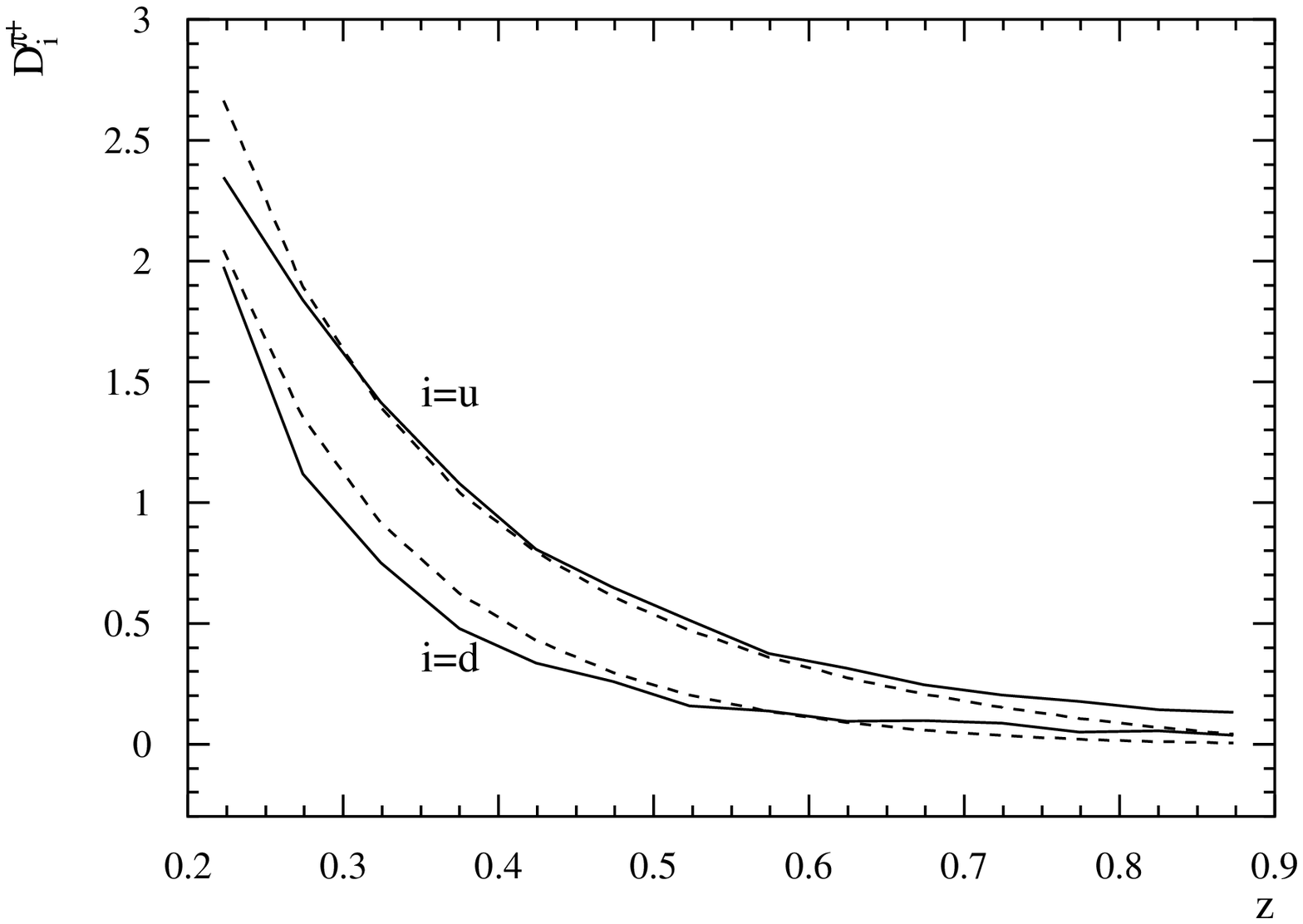,width=15cm}
\vspace{-1cm}
\caption{
A comparison of the extracted FFs
$D_u^{\pi^+}$, $D_d^{\pi^+}$ (solid lines)
at $\left< Q^2 \right> = 2.5\ {\rm GeV}^2$ as in 
Fig.~\ref{fig1} to the LO-parametrization in \cite{Kr} 
(dashed) where 
$(1-z) D_u^{\pi^+}(z) = D_d^{\pi^+}(z)$ at the input $Q_0^2$. 
}
\label{fig3}
\end{figure}
}

\def\figd
{
\begin{figure}[p]
\hspace{-2cm}
\epsfig{figure=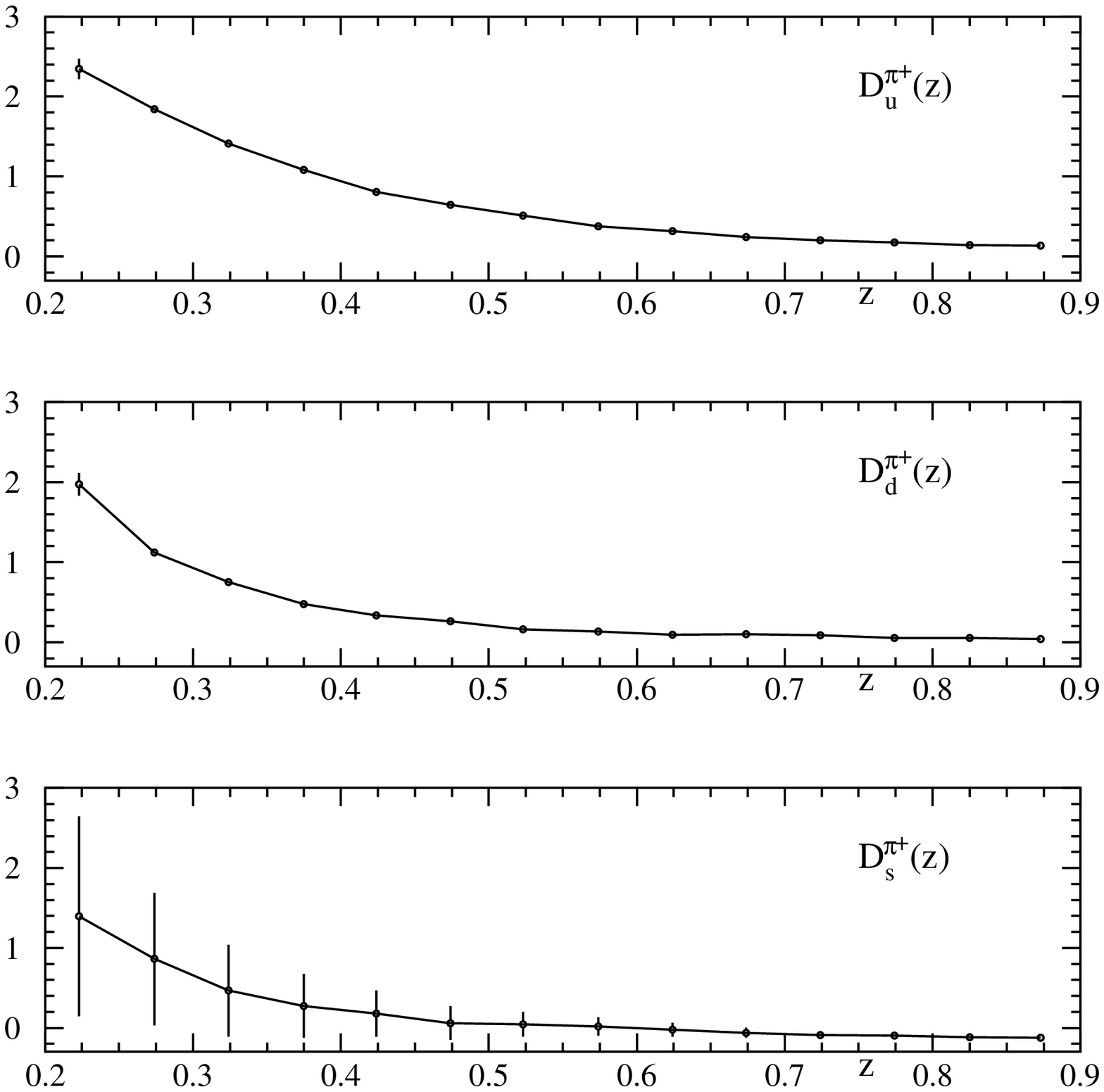,width=18cm}
\vspace{-1cm}
\caption{
The extracted fragmentation functions with errors which combine 
the experimental errors from \cite{HERMES} with a typical
20\% uncertainty
arising
from the evolution of the singlet fragmentation function.
}
\label{fig4}
\end{figure}
}

\def\fige
{
\begin{figure}[t]
\vspace{-1cm}
\hspace{2cm}
\epsfig{figure=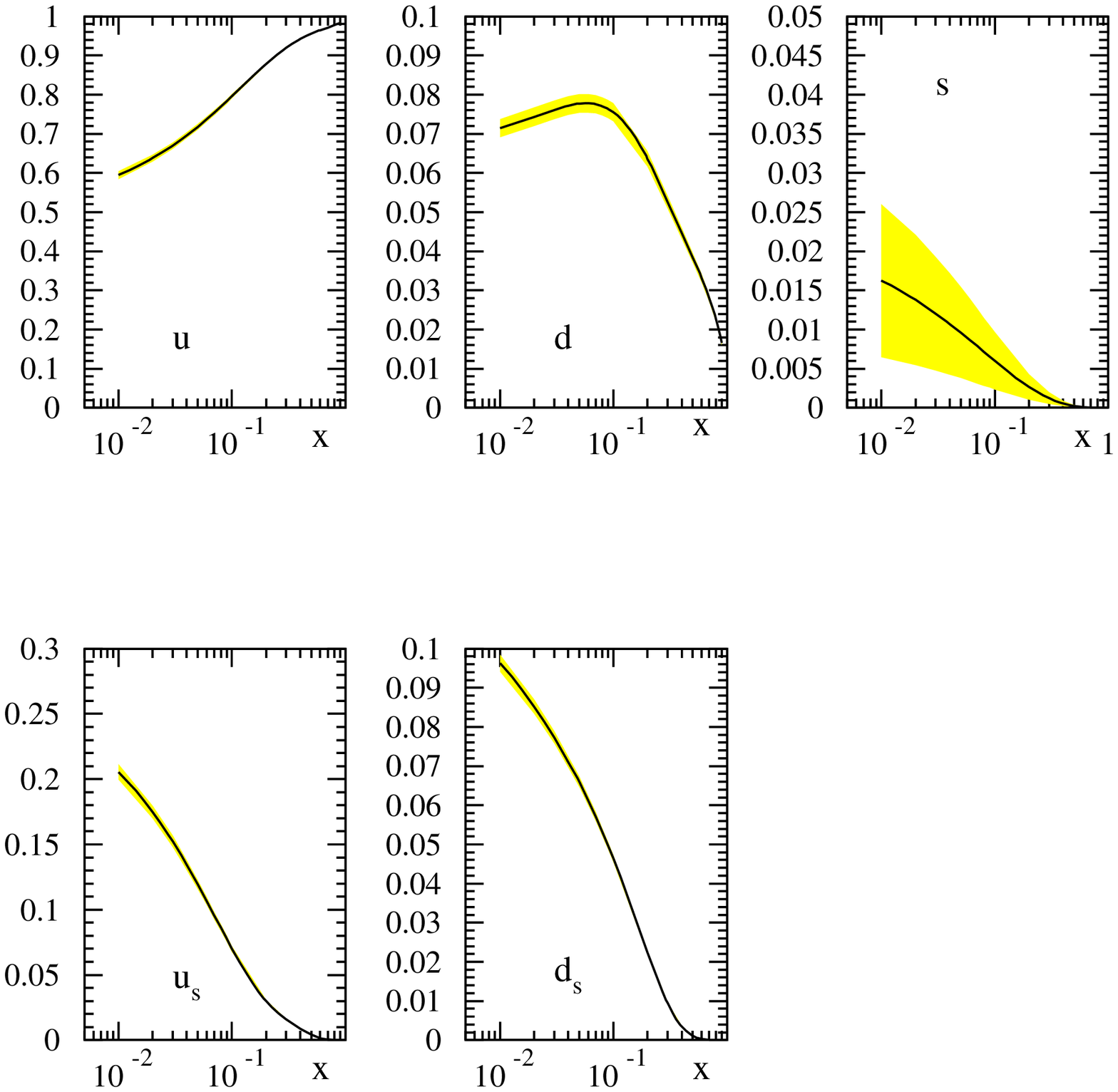,width=12cm}
\vspace{-1cm}
\caption{
Purity functions $P_{q/p}^{\pi^+}(x)$ extracted from HERMES data
\cite{HERMES01} with 20\% uncertainty from the evolution of the 
flavour-singlet $D_\Sigma^\pi$.
}
\label{fig5}
\end{figure}
}

\def\figf
{
\begin{figure}[t]
\vspace{-1cm}
\hspace{2cm}
\epsfig{figure=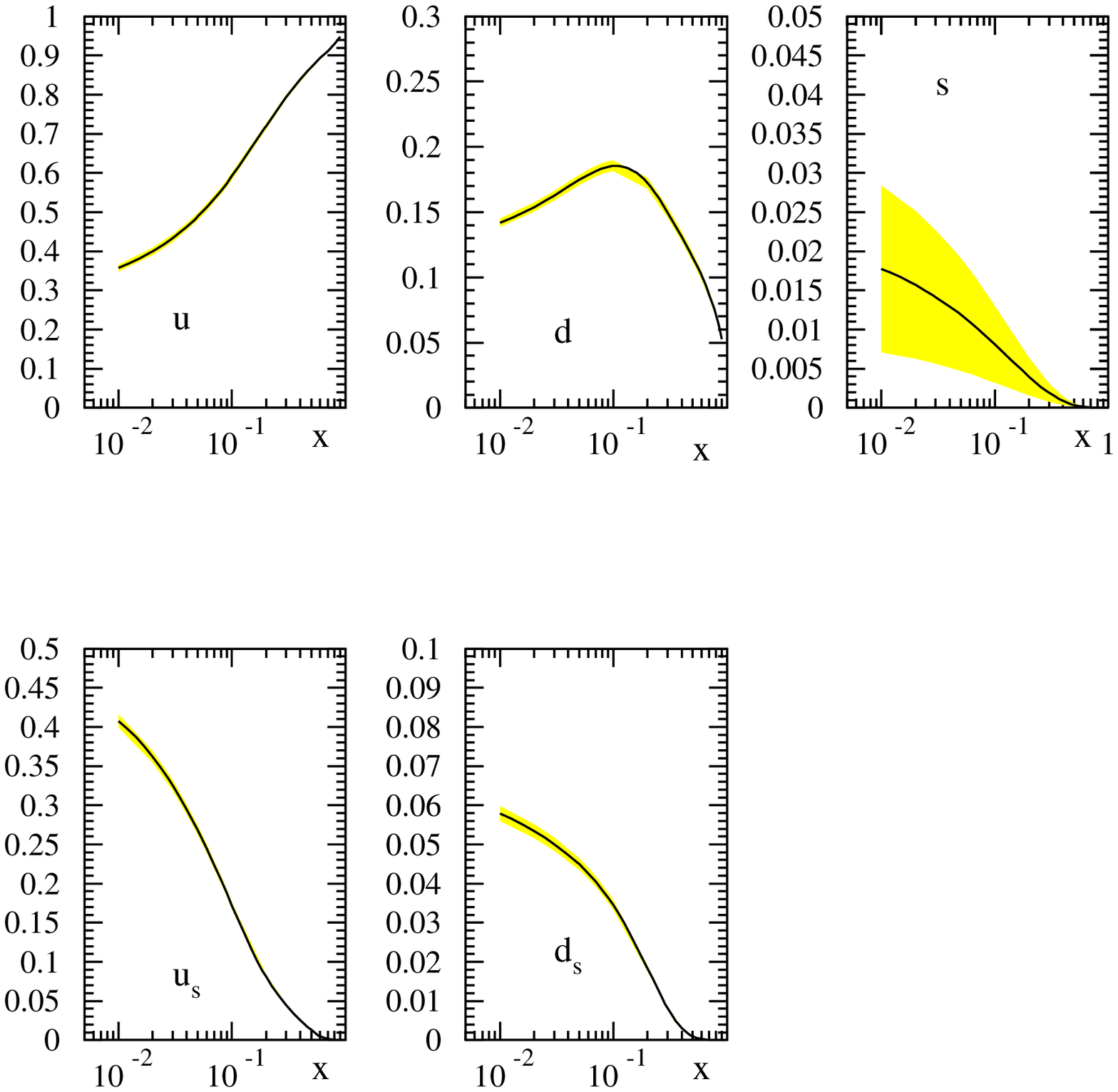,width=12cm}
\vspace{-1cm}
\caption{
Purity functions $P_{q/p}^{\pi^-}(x)$ extracted from HERMES data
\cite{HERMES01} with 20\% uncertainty from the evolution of the 
flavour-singlet $D_\Sigma^\pi$.
}
\label{fig6}
\end{figure}
}

\def\figg
{
\begin{figure}[p]
\vspace{-2cm}
\hspace{-1cm}
\epsfig{figure=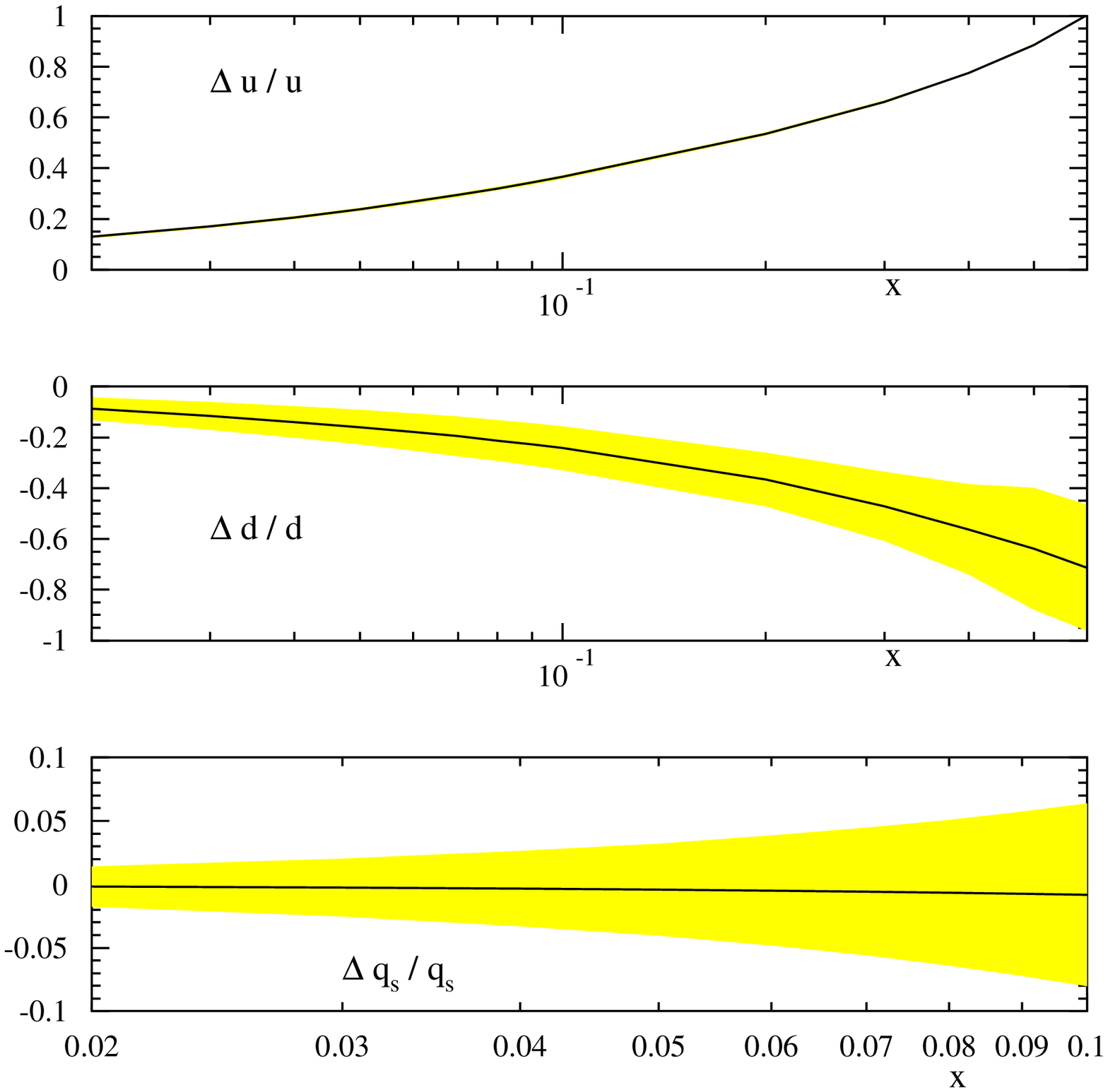,width=18cm}
\vspace{-1cm}
\caption{
Polarized parton densities extracted from the perfect
"data" as explained in the text. The HERMES simplifying assumption
$\Delta\bar u/\bar u=\Delta\bar d/\bar d=\Delta s/ s = \Delta q_s/ q_s$
has been used.
The uncertainties arise solely from the realistic experimental errors
on our fragmentation functions.
}
\label{fig7}
\end{figure}
}

\begin{document}

\begin{titlepage}
\begin{flushright}
MSU-HEP-10702
\end{flushright}

\begin{center}
{\large\bf  Fragmentation functions from semi-inclusive DIS pion production
and implications for the polarized parton densities}

\vspace{1cm}
{\bf Stefan Kretzer}\\
{\it Department of Physics and Astronomy, 
Michigan State
University, East Lansing,\\ e-mail: kretzer@pa.msu.edu}\\
{\bf Elliot Leader} \\{\it High Energy Physics Group, Imperial
College, London,\\ e-mail:
e.leader@ic.ac.uk}\\
 {\bf Ekaterina Christova}\\
{\it Institute of Nuclear Research and Nuclear Energy,
Sofia,\\
 e-mail: echristo@inrne.bas.bg}
\end{center}
\vspace{0.5cm}
\begin{abstract}

By combining 
recent  HERMES data on 
semi-inclusive DIS (SIDIS)
$\pi^\pm$-production with the singlet fragmentation function
$D_\Sigma^{\pi^+}$, which is well determined from
$e^+e^-$ data,  we are able to extract, for the first time, the
flavoured fragmentation functions  $D_u^{\pi^+}$,
 $D_d^{\pi^+}$ and $D_s^{\pi^+}$ 
without making any assumptions about
  favoured and unfavoured transitions.
  Whereas $D_u^{\pi^+}$ and  $D_d^{\pi^+}$ are very well determined, the
accuracy of $D_s^{\pi^+}$ is limited by the uncertainty in
evolving $D_\Sigma^{\pi^+}$ 
from the $Z^0$ pole down to the SIDIS scale of
a few $(GeV)^2$. We discuss how the precision on $D_s^{\pi^+}$
could be improved.
Knowledge of the 
$D_{q=u,d,s}^{\pi^+}$ will permit the extraction of the
polarized parton densities from future polarized SIDIS asymmetry
measurements. We study the precision that can be expected in such
an extraction.

\end{abstract}
\end{titlepage}
\newpage


\newpage
\setcounter{page}{1}
\section{Introduction \label{sect:intro}}

Fully inclusive deep inelastic scattering (DIS) of the neutral
type
\be
l^\pm + N \to l^\pm + X
\ee
yields information only on
the combination of parton densities $q(x) + \bar q(x)$, in
the unpolarized case, and on the combination of polarized parton
densities  $\Delta q(x) + \Delta \bar q(x)$ when longitudinally polarized leptons
interact with longitudinally polarized nucleons. It is crucially from reactions with
neutrinos and antineutrinos, in the unpolarized case, that a separate knowledge
of the parton $q(x)$ and antiparton $\bar q(x)$ densities can be inferred.
 This avenue is, at present, not open to the polarized case.

 The main approach to a separate knowledge of the $\Delta q(x)$ and $\Delta \bar q(x)$
thus rests upon the growing activity in the field of polarized semi-inclusive
deep-inelastic experiments of the type
\be
\overrightarrow{l}^\pm + \overrightarrow{N}\to l^\pm +h+ X\label{eq:1}
\ee
where $h$ is the detected hadron.

The cross sections (or spin asymmetries) for such reactions
depend, in leading order QCD, upon products of parton densities
and fragmentation functions (FFs) $D_q^h(z)$ for a parton $q$ to
fragment into hadron h. (In NLO QCD these products become
convolutions.)

It has been shown \cite{we1,we2} that if systematic errors can be well
enough controlled so as to allow a meaningful combination of data
from different targets and hadrons of different charge, there is
sufficient information to extract information on both the
polarized parton densities and the fragmentation functions.

In the past this has not been possible and the strategy adopted in the analysis of
the experimental data \cite{SMC,HERMES,deFlorian} has been to assume a complete knowledge of
the unpolarized
 densities $q(x)$, $\bar q(x)$ and of the fragmentation functions $D_q^h(z)$, $D_{\bar q}^h(z)$.
With these, in \cite{HERMES} an auxiliary function was constructed, 
the flavour ($q=u,d,s,{\bar u},{\bar d},{\bar s}$) purity 
$P_{q/N}^h(x,z)$ \cite{purity}
for each hadron $h$ and for each target nucleon $N$. 
Given the purities,
the polarized data can then, in principle,
be used to directly extract the polarized densities $\Delta q(x)$,  $\Delta \bar q(x)$.

However, the situation has now changed because recently the HERMES
group has for the first time published unpolarized 
charge separated data for $\pi^\pm$ production on a proton target
\cite{HERMES01}.  The main aim of our paper is to demonstrate
that this data, taken in conjunction with the information on the
flavour singlet combination $D_{\Sigma}^{\pi^+}$ of FFs which can
be fairly reliably obtained from the data on $e^+e^- \to \pi^\pm
X$ at the $Z^0$ peak, allows a first direct determination of the
FFs $D_u^{\pi^+}$, $D_d^{\pi^+}$ and $D_s^{\pi^+}$.
We note the caveat that the data in \cite{HERMES01} 
covering $0.2 < z < 0.9$
exhibit
large ${\cal O}(40 \% )$ isospin violations at large 
$z \gtrsim 0.7$. 
It is most unlikely that such a large
breaking of isospin invariance can be a genuine effect
in the current fragmentation region of 
semi-inclusive DIS (SIDIS), and it would seem
unnatural to incorporate it 
into an FF formalism. 
A possible explanation for the effect is given in \cite{HERMES01}.
Our analysis will therefore be relevant
mainly for intermediate $0.2 < z \lesssim 0.7$.

It turns out that the least well determined FF is $D_s^{\pi^+}$, since
it is most dependent on the evolution downwards of
$D_{\Sigma}^{\pi^+}(z,Q^2)$ all the way from $Q^2\approx m_{Z^0}^2$
to $Q^2 \approx$ few $(GeV)^2$, and this involves mixing with the
gluon FF $D_G^{\pi^+}(z,Q^2)$. We try to assess the sort of
accuracy required in future $e^+e^-$ measurements in order to
achieve an accuracy of 20--30\% on $D_s^{\pi^+}(z)$.

In order to study the accuracy of the polarized parton densities
obtained in the past via the use of purity \cite{HERMES} 
we
proceed as follows. Firstly,  we use our FFs to calculate the
central value and errors of the 
integrated purity function $\int dz\ P_{q/p}^{\pi^+}(x,z)$
and $\int dz\ P_{q/p}^{\pi^-}(x,z)$ for a proton target. 
Then, because at present the polarized SIDIS
data does not exist separately for $\pi^+$ and $\pi^-$, we take
the central values of the published polarized
parton densities of Leader,
Sidorov and Stamenov \cite{polDIS} obtained purely from DIS data,
and using the central values of the purity we generate fake
"data" on polarized SIDIS asymmetries $\Delta A_p^{\pi^+}$ and
$\Delta A_p^{\pi^-}$,
  and also on the polarized DIS asymmetries. We
then go through 
a similar procedure as adopted by the HERMES group to
obtain from this "data" the polarized parton densities, with this
difference, that we allow for the uncertainty in the value of the
purity function. In this way we obtain an indication of the
uncertainty in the polarized parton densities inherent in the
purity approach. 

In this paper we work to leading order (LO) in QCD as the
purity concept only makes sense in LO and because in LO we can
deal with simple algebraic equations which are 
physically most transparent to interpret 
and have a well-defined error propagation.
Of course, in the
long run, a more complete NLO analysis will be required.
Standard experimental techniques \cite{HERMES} are based on an {\it ad hoc}
combination of LO (polarized) parton distributions
with e.g.~LUND-type Monte Carlo fragmentation functions. 
Such an effective approach cannot be extended to NLO without
a highly non-trivial definition of the long- and short distance pieces
in the MC environment which is -- to our knowledge -- lacking
at present.
It will, therefore, be vital
to bring the measurements
in touch with well-defined  factorized\footnote{The 
factorization of $x-$ and $z-$dependence at LO is
an artefact from the one particle phase space 
(delta-function) of LO-SIDIS and is of no 
fundamental importance as opposed to the mass factorization
of NLO cross sections.} 
QCD approaches \cite{nlo} 
combining universal 
parton distribution functions
(PDFs) with universal FFs
because, otherwise, the extracted PDFs will not have
any physical relevance. The most important experimental 
information will 
be on scheme- and model-independent data for cross sections
and not on the extracted (unobservable) PDFs and FFs.


\section{Extraction of Fragmentation Functions \label{sect:FF}}
\subsection{Formalism}
For a leading order treatment we follow the notation of \cite{we2}
and remove some kinematical factors by introducing for the DIS
 and SIDIS cross sections on a proton target:
 \be
\tilde \sigma^{DIS}
&=& \frac{x(P+l)^2}{4\pi
\alpha^2}\left (\frac{2y^2}{1+(1-y)^2}\right )
\frac{d^2\sigma^{DIS}}{dx\,dy}\label{sigmaDIS}\\
\Delta \tilde \sigma^{DIS} &=&  \frac{x(P+l)^2}{4\pi
\alpha^2}\left (\frac{y}{2-y}\right ) \left
[\frac{d^2\sigma_{++}^{DIS}}{dx\,dy} -
\frac{d^2\sigma_{+-}^{DIS}}{dx\,dy}\right ]\\
\tilde \sigma^h &=&  \frac{x(P+l)^2}{4\pi \alpha^2}\left
(\frac{2y^2}{1+(1-y)^2}\right )
\frac{d^3\sigma^h}{dx\,dy\,dz}\label{sigmah}\\
\Delta
\tilde \sigma^h &=&  \frac{x(P+l)^2}{4\pi \alpha^2}\left
(\frac{y}{2-y}\right ) \left [\frac{d^3\sigma_{++}^h}{dx\,dy\,dz}
- \frac{d^3\sigma_{+-}^h}{dx\,dy\,dz}\right ]\label{Deltasigma}
\ee
Here,
$P^\mu$ and $l^\mu$ are the nucleon and
lepton four momenta,  and
$\sigma_{\lambda\nu}$ refers to a lepton of
helicity $\lambda$ and a nucleon of helicity $\nu$. 
The variables $x,y,z$ are the usual DIS kinematic variables.
Then one has the very simple LO results:
\be
\tilde\sigma^{DIS} (x,Q^2)& =&
\sum_{q,\bar q} e^2_q \,q_i(x,Q^2)\,\\
 \Delta \tilde\sigma^{DIS}
(x,Q^2) &=&  \sum_{q,\bar q} e^2_q
\,\Delta q_i(x,Q^2)\\
\Delta \tilde \sigma^h (x,z,Q^2) &=& \sum_{q,\bar q} e^2_q\,
\Delta q_i(x,Q^2)\,
D_i^h(z,Q^2) \\
\tilde \sigma^h (x,z,Q^2) &=&  \sum_{q,\bar q} e^2_q
\,q_i(x,Q^2)\, D_i^h(z,Q^2),\label{eq:sigmaLO}
 \ee

Note that the inclusion of a factor 
$(1+R)/(1+\gamma^2)$
in (\ref{eq:sigmaLO}) (see e.g.\ Eq.~(5) of \cite{HERMES}) is not
justified theoretically. The correct handling of the longitudinal
cross-section is a more complicated NLO effect in SIDIS
(see Eqs.~(56) - (60) of \cite{we2}). Here, as mentioned, we work to LO
only.
Specializing now to $\pi^\pm$ production we introduce 
the measured observables\footnote{
As we are considering positive and negative
charges seperately we are {\it not} adopting the convention
$h^\pm \equiv h^+ + h^-$ \cite{Binnewies,Kr,KKP}.
}
\be
R_p^{\pi^\pm }(x,z,Q^2)& \equiv& \frac {\sigma_p^{\pi^\pm}}
{\sigma^{DIS}_p}\label{eq:Rp1}\\
& =&
 \frac {\tilde\sigma_p^{\pi^\pm}}
{\tilde\sigma^{DIS}_p}\label{eq:Rp2}
 \ee
 Using charge conjugation
and isospin invariance we require only 3 independent FFs:
\be
D_u^{\pi^+}(z,Q^2),\qquad D_d^{\pi^+}(z,Q^2),\qquad
D_s^{\pi^+}(z,Q^2) \label{eq:DC}
\ee
The remaining ones are then:
\be
D_{\bar u}^{\pi^-}=D_d^{\pi^-}=D_{\bar d}^{\pi^+}=D_u^{\pi^+}\label{eq:SU1}\\
D_{\bar d}^{\pi^-}=D_u^{\pi^-}=D_{\bar u}^{\pi^+}=D_d^{\pi^+}\label{eq:SU2}\\
D_{\bar s}^{\pi^-}=D_s^{\pi^-}=D_{\bar s}^{\pi^+}=D_s^{\pi^+}\label{eq:SU3}
 \ee
 Thus
 \be R_p^{\pi^+}
&=&\frac{1}{\tilde\sigma^{DIS}_p}\left\{\frac{4}{9}\left(uD_u^{\pi^+}
+\bar u D_{\bar u}^{\pi^+}\right)+\frac{1}{9}\left(dD_d^{\pi^+} +
\bar dD_{\bar d}^{\pi^+} +
sD_s^{\pi^+} + \bar s D_{\bar s}^{\pi^+}\right)\right\}\nn\\
&=& \frac{1}{9\tilde\sigma^{DIS}_p}\left\{(4 u+\bar d)D_u^{\pi^+}
+( 4\bar u +d)D_d^{\pi^+} + \left( s + \bar
s\right)D_s^{\pi^+}\right\}\label{eq:Rp+}
 \ee
 Similarly
 \be
 R_p^{\pi^-} =
\frac{1}{9\tilde\sigma^{DIS}_p}\left\{(4 \bar u+ d)D_u^{\pi^+} +(
4 u +\bar d)D_d^{\pi^+} + \left( s + \bar
s\right)D_s^{\pi^+}\right\}\label{eq:Rp-}
\ee
Assuming a good
knowledge of the unpolarized parton densities we can immediately
obtain
\be
D_u^{\pi^+}- D_d^{\pi^+} =
\frac{9\left(R_p^{\pi^+} -
R_p^{\pi^-}\right)\tilde{\sigma}_p^{DIS}}{4u_V -d_V }\label{eq:D-}
 \ee

 In order to obtain $D_u^{\pi^+}+ D_d^{\pi^+} $ and $D_s^{\pi^+}$
 we require one further
 piece of experimental information. We shall argue that  it can
 be obtained from the data on 
$e^+e^-\to \pi^{\pm} X$ at the $Z^0$
 peak.

 \subsection{Use of the $e^+e^-$ data}

For some time it was believed that the fragmentation functions
obtained by 
Binnewies 
et al.\ \cite{Binnewies}, from a detailed analysis of
the $e^+e^-$ data over a wide range of energies, were reasonably
well determined. 
However, recent analyses \cite{Kr,KKP,BFGW} have shown
that equally good fits to $e^+e^-$ data can be achieved with FFs
of a given flavour which differ widely from each other. The
$e^+e^-$ data do not, therefore, constrain the FFs of a given
flavour very well, and, in retrospect, this is really not
surprising.

However, by a piece of good fortune, the $e^+e^-$ data at the
$Z^0$ peak directly 
measure a linear combination of FFs which is
very close to the 
${\rm SU(3)_f}$ flavour singlet combination, i.e. in
 \be
\label{eq:meas}
D_{\rm meas}^{\pi^+ + \pi^-}&=& 
\sum_{q=u,d,s} \left( D_q^{\pi^+ + \pi^-} + 
D_{\bar q}^{\pi^+ + \pi^-}\right) 
{\hat e}_q^2(s)
\ee 
the squared electroweak couplings 
$ {\hat e}_q^2(s) $ from ${\rm SU}(2) \times {\rm U}(1)$
gauge symmetry (given e.g.\ in the 
appendix of \cite{Kr}) are
flavour-independent to within $\sim$ 25 \% at $\sqrt{s}=M_Z$ as
opposed to a relative factor of 4 for the electromagnetic
couplings of up- and down-type quarks at lower cms energies. 
The exact singlet below in (\ref{eq:Dsing}) 
would correspond to a measurement
at an $e^+ e^-$ cms energy of 
$\sqrt{s}=78.4\ {\rm GeV}$ or $\sqrt{s}=113.1\ {\rm GeV}$
where it happens that 
${\hat e}_u^2(s) = {\hat e}_d^2(s)= {\hat e}_s^2(s)  $. 
Accordingly (\ref{eq:meas}) is approximately proportional
to the singlet combination  
 \be 
D_{\Sigma}^{\pi^+}&\equiv& \left(D_u^{\pi^+}+
 D_{\bar u}^{\pi^+} + D_d^{\pi^+}+ D_{\bar d}^{\pi^+} + D_s^{\pi^+}+
D_{\bar s}^{\pi^+} \right)\\
&=&
2\left( D_u^{\pi^+}+  D_d^{\pi^+}+  D_s^{\pi^+}\right)\label{eq:Dsing}
\ee
where we have used charge conjugation and
eqs.\eq{SU1} - \eq{SU3} in the last step.
Using isospin and charge conjugation
invariance $D_u^{\pi^\pm} + D_{\bar u}^{\pi^\pm}
= D_d^{\pi^\pm} + D_{\bar d}^{\pi^\pm}$
and approximating 
$\left. {\hat e}_u^2(s) / {\hat e}_d^2(s) \right|_{s=M_Z^2}
=  \left.{\hat e}_u^2(s) / {\hat e}_s^2(s) \right|_{s=M_Z^2}
\simeq 3 / 4$ we can write the singlet combination
\be
\label{eq:approx1}
D_{\Sigma}^{\pi^+} =
\frac{4}{7} {\tilde D}_{\rm meas}^{\pi^+ + \pi^-}
-\frac{1}{7}\left(
D_s^{\pi^+}+D_{\bar s}^{\pi^+} \right)
\ee
where we have introduced a convenient change in normalization
\be
\label{eq:norm}
{\tilde D}_{\rm meas}^{\pi^+ + \pi^-} = D_{\rm meas}^{\pi^+ + \pi^-}
\ /\ {\hat e}_d^2(s)
\ee
The {\it extreme} limits  
$0 < (D_s^{\pi^+}+D_{\bar s}^{\pi^+}) < 
(D_u^{\pi^+} + D_{\bar u}^{\pi^+})$ 
then correspond to
\be \label{eq:approx2}
\frac{4}{7}\ {\tilde D}_{\rm meas}^{\pi^+ + \pi^-} < 
D_{\Sigma}^{\pi^+}
< \frac{6}{11}\ {\tilde D}_{\rm meas}^{\pi^+ + \pi^-}
\ee
i.e.\ to only a $\sim 5\%$ uncertainty for $D_{\Sigma}^{\pi^+}$.

Not surprisingly, therefore, the singlet FFs in the analyses
\cite{Kr,KKP,BFGW} agree 
with each other to better than 5\% for $ 0.2 < z < 0.7 $
as seen in Fig.~\ref{fig1}.
\figa
We may thus take as a known quantity $D_{\Sigma}^{\pi^+}(z,Q^2=m^2_{Z^0})$
and from Fig.~\ref{fig1} we observe a stable evolution down to 
$Q^2 = 100\ {\rm GeV}^2$.

But we require this quantity at a scale of a few $({\rm GeV})^2$ and it
 thus has to be evolved down through a large range of $Q^2$, and in this
 evolution mixes with the poorly known gluon FF $D_G^{\pi^+}$. (Of course
 we cannot carry out
the evolution of $D_{\rm meas}$ itself since it
 is a combination of  singlet and  non-singlet pieces and we do
 not know the values of these separately.)
The FF analyses in \cite{Binnewies,KKP,Kr,BFGW} cover data down to
$\sqrt{s} \simeq 30\ {\rm GeV}$ and from Fig.~\ref{fig1} we judge
this fixes a stable singlet FF down to $\sqrt{s} \simeq 10\ {\rm GeV}$.
Below, however, the evolution uncertainties set in and from the right
of Fig.~\ref{fig1} we quantify this uncertainty conservatively to be 
a $\sim 20\ \%$ effect uniformly in $z$. We convinced ourselves this
is indeed a typical order of magnitude by comparing the several sets
of LO and NLO FFs for $\pi^+ + \pi^- , K^+ + K^- , h^+ + h^- $ in 
\cite{Binnewies,KKP,Kr,BFGW} and not only the two sets plotted 
in Fig.~\ref{fig1}. Clearly, a low scale measurement of the singlet
FF or a resolution of the evolution ambiguities through a
determination of the gluon FF would be highly desirable information.

Subject therefore to possible errors due to the evolution, we have available
the additional experimental data that we require, and we then obtain:
\be
D_u^{\pi^+}+  D_d^{\pi^+}=\frac{9\left(R_p^{\pi^+} +
 R_p^{\pi^-}\right)\tilde\sigma_p^{DIS} -2s\,D_{\Sigma}^{\pi^+}}{4(u+\bar u -s) +d+\bar d}
 \label{eq:D+}
\ee
and
\be
D_s^{\pi^+}=\frac{-18\left(R_p^{\pi^+} +
 R_p^{\pi^+-}\right)\tilde\sigma_p^{DIS} +[4(u+\bar u) +d+\bar d]\,D_{\Sigma}^{\pi}}
 {2\,[4(u+\bar u -s) +d+\bar d\,]}
 \label{eq:Ds}
\ee
We note that the singlet FF plays no role in eq.~(\ref{eq:D-})
and that its weight increases in going from (\ref{eq:D+}), where it is multiplied
by the suppressed strange PDF, to (\ref{eq:Ds}), where it is multiplied by the
unsuppressed $u$ and $d$ PDFs.
Correspondingly, 
we must expect an increasing importance of the
propagation of the $\sim 20 \%$ error of
$D_{\Sigma}^{\pi^+}$ into these equations.

The LH sides of eqs. \eq {D-},\eq{D+} and \eq{Ds} are functions of $z$ and $Q^2$,
whereas the RH sides are, in principle, functions of $x,z$ and $Q^2$. Only in the LO
approximation does the variable $x$ become a 
{\it passive} variable \cite{we2} i.e.
there is no dependence on it. Strictly one should test for this lack of $x$-dependence,
as a measure of the reliability of the LO treatment. However, in this paper, in order
to improve statistics, we shall take it for granted that the LO treatment
is adequate.

\subsection{Combined analysis of HERMES and $e^+e^-$ data}

The formalism given in \eq{SU1}-\eq{SU3}, 
\eq{D-}, \eq{Dsing}, \eq{D+} and \eq{Ds}
presupposes the availability of data at fixed $x$ and $y$. In fact
the available HERMES data is integrated over the kinematic range \cite{HERMES}
$Q^2 > 1\ ({\rm GeV/c})^2,\ W^2 > 10\ {\rm GeV}^2,\ y < 0.85$.
Handling integrated data slightly complicates the formalism since
the $y$-dependent factors in the numerator and denominator  of
\eq{Rp1} no longer cancel out to give the simpler result
\eq{Rp2}. We have done the analysis using the data integrated
over the kinematic range of the experiment and  have checked that
using the simpler formalism 
with \cite{HERMES}
\be
&&x=<x>=0.082\\
&&Q^2=<Q^2>=2.5 \,({\rm GeV/c})^2 \\
&&W^2=<W^2>= 28.6 \,({\rm GeV})^2
\ee
makes no discernible difference to the results for the FFs.

The stability of our results for the central values
of $D_u^{\pi^+}$, $D_d^{\pi^+}$
is studied in Fig.~\ref{fig2}.
\figb
The NLO determination of
the singlet combination $D_\Sigma^{\pi^+}$ due to
Kretzer~\cite{Kr} was utilized. To test the stability of our results
we have used two different sets of unpolarized parton densities.
We found the effect of employing, respectively, the NLO
MRST~\cite{MRST} or the NLO GRV~\cite{GRV} unpolarized parton
densities in eqs. \eq{Rp+}, \eq{Rp-}, \eq{D-}, \eq{D+} and \eq{Ds}
leads to a negligible $\sim 5 \%$ effect.
This is because the unpolarized
densities are very well constrained in the region of interest. We
have checked that use of the LO GRV densities also have almost no
noticeable effect.

The FFs  $D_u^{\pi^+}$ and  $D_d^{\pi^+}$ are quite well
constrained by the SIDIS data, but, as expected, $D_s^{\pi^+}$ is
sensitive to the singlet combination of FFs determined
 from $e^+e^-$ data. This can be clearly seen in Fig.~\ref{fig2} where we
 compare the results for each $D_q^{\pi^+}$ using 
 NLO versions of $D_\Sigma  ^{\pi^+}$ as obtained by Kretzer
 \cite{Kr} 
and Kniehl, Kramer and P\"{o}tter \cite{KKP}
from the $e^+e^-$ data. As mentioned in the
 Introduction, $D_\Sigma  ^{\pi^+}$ is very well determined at the
 $Z^0$ peak, but the mixing, under evolution, with the gluon FF
 $D_G^{\pi^+}$ induces an uncertainty of about $10-20\%$ at
 $Q^2=2.5$
 $({\rm GeV/c})^2$.
As can be seen in Fig.~\ref{fig2}, $D_s^{\pi^+}$ may
even turn unphysically negative at large $z \gtrsim 0.7$ where
our analysis is not supposed to be reliable, anyway, as mentioned
in the Introduction.
We note Fig.~\ref{fig2} shows the typical effect expected from
the uncertainty of $D_\Sigma  ^{\pi^+}$ and that a 
similar picture emerges if we switch 
$D_\Sigma  ^{\pi^+}$ between the LO and NLO parametrizations of
\cite{Kr}: The stability of $D_u^{\pi^+}$ and  $D_d^{\pi^+}$ 
is remarkable. $D_s^{\pi^+}$, on the other hand,
changes significantly. If a measurement could fix
$D_\Sigma  ^{\pi^+}(Q^2\simeq 2.5\ {\rm GeV}^2)$ to within
$\sim 5 \%$ we would have a handle on $D_s^{\pi^+}$ as well
at the $\sim 20-30 \%$ level. 

 In summary we see that $D_u^{\pi^+}$ and  $D_d^{\pi^+}$ are
 remarkably well constrained by the SIDIS data. $D_s^{\pi^+}$
 however, is undetermined within a factor of about 2. 
In Fig.~\ref{fig4}
\figd
we show the final results for
our FFs and our estimate of their errors. 
These include Gaussian error propagation of the
experimental errors in \cite{HERMES} combined with
a 20 \% error of $D_\Sigma  ^{\pi^+}(Q^2= 2.5\ {\rm GeV}^2)$.
The FFs can be
 described analytically\footnote{
Employing $D_q^{\pi^+}=N\ z^{\alpha}\ (1 - z)^{\beta_q}$
ans\"{a}tze with flavour independent $N,\alpha$ at 
$\left< Q^2 \right> =2.5 {\rm GeV/c}^2$ slightly
worsens the quality of the parametrization.}
at $\left< Q^2 \right> =2.5 {\rm GeV/c}^2$ by
 \be
D_u^{\pi^+}=0.689\ z^{-1.039}\ (1 - z)^{1.241}
\\
D_d^{\pi^+}= 0.217\ z^{-1.805}\ (1 - z)^{2.037}
\\
D_s^{\pi^+}=0.164\ z^{-1.927}\ (1 - z)^{2.886}
\ee
where $D_s^{\pi^+}$ has to be taken with care as Fig.~\ref{fig4} 
shows that basically no value within
$0<D_s^{\pi^+}<D_u^{\pi^+}$ can be excluded at present.

 Finally, as a matter of interest, we compare in 
Fig.~\ref{fig3} 
\figc
the $D_{q=u,d,s}^{\pi^+}$
 obtained in this paper with the LO $D_q^{\pi^+}$ obtained by
 Kretzer~\cite{Kr} purely from an analysis of the $e^+e^-$
 data. In the latter the flavour separation is not fixed by the
 data and is somewhat {\it ad hoc} and assumed $D_d^{\pi^+}=
 D_s^{\pi^+}$ and $D_u^{\pi^+} >D_d^{\pi^+}$ 
by imposing $(1-z)\ D_u^{\pi^+} = D_d^{\pi^+}$ 
at the input scale. 
Surprisingly,
 Kretzer's $D_q^{\pi^+}$ are not very different from the
 $D_q^{\pi^+}$ obtained from our analysis of the SIDIS data!

\section{Implications for the polarized parton densities}

As mentioned in the Introduction, the absence of neutrino data
for polarized DIS means that the extraction of the individual
$\Delta q(x)$ is impossible. Only the combination $\Delta q(x) +
\Delta \bar q(x) $ can be found and the 
flavour separation of
these relies heavily on the evolution in $Q^2$ and is thus
unreliable, given the small range of $Q^2$ available in polarized
DIS experiments. Thus polarized SIDIS has a vital role to play in
this matter.

At present, however, there are no published data for polarized
$\pi^\pm$-production, though there does exist data for
undifferentiated polarized $h^\pm$- production, which have been
used by the HERMES group to extract information on the polarized
parton densities via what is known as the purity method.

We believe that in this approach the errors on the polarized
parton densities 
are somewhat underestimated 
and we shall use our FFs
to study this question.

 The flavour $q$ purity function for
 protons\cite{purity} used by the HERMES
 group\cite{HERMES}\footnote{Note that the graphs shown
 in the HERMES publications~\cite{Confs} and labelled ``purity'' are
actually plots of an ``effective purity''
 incorporating various experimental cuts.} is defined by
 \be
 P_{q/p}^{h}(x)
 =\frac{e^2_q\,q(x)\,\int_{0.2}^1D_q^{h}(z)\,dz}
 {\sum_{q'}e^2_{q'}\,q'(x)\,\int_{0.2}^1D_{q'}^{h}(z)\,dz}
 \label{eq:Pf}
 \ee
 where again, we utilize the MRST parton densities and take $Q^2 =
 <Q^2>$.

 Defining now the SIDIS spin asymmetry
 \be
 <\Delta A_p^{h}(x)> \equiv \frac{\int_{0.2}^1dz
 \,\Delta\tilde\sigma_p^{h}(x,z)}{\int_{0.2}^1dz\,
 \tilde\sigma_p^{h}(x,z)}\label{eq:deltaAp}
 \ee
  we have in LO,
 \be
 <\Delta A_p^{h}(x)>=\sum_q\,P_{q/p}^{h}(x) \left(\frac{\Delta
 q(x)}{q(x)}\right).\label{eq:deltaApLO}
 \ee
 Similarly for the DIS spin asymmetry we can define
 \be
P_{q/p}^{DIS}(x)
 =\frac{e^2_q\,q(x)}
 {\sum_{q'}e^2_{q'}\,q'(x)}
 \label{PfDIS}
 \ee
 and then, in LO, we have, with $Q^2 =
 <Q^2>$:
 \be
\frac{
 \Delta\tilde\sigma_p^{DIS}(x)}{
 \tilde\sigma_p^{DIS}(x)}= \sum_q\,P_{q/p}^{DIS}(x) \left(\frac{\Delta
 q(x)}{q(x)}\right).\label{eq:deltaDIS}
 \ee
 Similar expressions for $<\Delta
 A_n^{h}(x)>$ and $\Delta\tilde\sigma_n^{DIS}(x)/
 \tilde\sigma_n^{DIS}(x)$ can be obtained in an obvious way.

 At each value of $x$ there are in principle 6 pieces of data 
($h=\pi^\pm$
 for $p$, $h=\pi^\pm$ for $n$, and DIS for $p$, $n$), so that there
 is enough information to {\it solve} for the 6 quark polarized
 densities $\Delta q(x)/q(x)$, for $q=u,\bar u, d, \bar d, s,
 \bar s$.  In the published analyses of the latter data
 ~\cite{HERMES} the HERMES group has preferred  to model the
 polarized sea with assumptions such as
 \be
 \frac{\Delta\bar u}{\bar u}=\frac{\Delta\bar d}{\bar d}=
 \frac{\Delta s}{ s}=\frac{\Delta\bar s} {\bar s}\equiv
 \frac{\Delta q_s}{q_s}\label{eq:sea1}
 \ee
 or
 \be
 \Delta\bar u =\Delta\bar d =\Delta s =\Delta\bar s
 \equiv \Delta q_s
 \label{eq:sea2}
 \ee
 and then to obtain the 3 independent polarized densities by
 making a best fit to the 6 -pieces of data at each $x$.

 The problem with this approach is that the purity functions were
 constructed using LUND model information on the FFs. 
We think \cite{we2}
this is an un-reliable procedure 
since a combination of (polarized) PDFs and 
LUND-type of FFs is at present lacking a rigorous
theoretical framework as opposed to 
our combination of universal (polarized) PDFs with universal
FFs in line with the
factorization theorems of QCD \cite{fact}.

 We argue that the above
approach much underestimates the uncertainty on
 the polarized parton densities. To illustrate this we construct purity functions and
 their errors for pion production, using the fragmentation functions determined
  by us and the unpolarized MRST parton densities. The formulae are exactly as
  in \eq{Pf}, \eq{deltaAp} and \eq{deltaApLO} with $h$ replaced by $\pi^+$. 
We show our calculated purities for protons
with errors in Figs.~\ref{fig5}, \ref{fig6}.
\fige
\figf
The size of the errors are as to be anticipated from the errors on
the $D_{q=u,d,s}^{\pi^+}$ in Fig.~\ref{fig4} and the definition
of the purities in Eq.~(\ref{eq:Pf}).
In the
 absence of separate $\pi^\pm$ SIDIS spin asymmetry data, we take
 the central values of the polarized parton densities, as derived
 by Leader, Sidorov and Stamenov~\cite{polDIS} from purely DIS data
  and by feeding these $\Delta q$
 into eqs. \eq{deltaApLO} and \eq{deltaDIS} and into the analogous one
 for $\Delta A_p^{\pi^-}$, in which we utilize  the central values of the
 purity functions, we generate a set of fake "data" for
 $\Delta A_p^{\pi^+}$, $\Delta A_p^{\pi^-}$ and
 $\Delta\tilde\sigma_p^{DIS}/\tilde\sigma_p^{DIS}$.

 Having now this set of fake ``data'' we forget where it came from and use
 it to  solve for the polarized parton densities, mimicking the approach used
 by the HERMES group. Thus we take
\be
 \frac{\Delta\bar u}{\bar u}=\frac{\Delta\bar d}{\bar d}=
 \frac{\Delta\bar s} {\bar s}\equiv\frac{\Delta q_s}{ q_s}\label{eq:sea3}
 \ee
 and solve \eq{deltaDIS}, \eq{deltaAp} 
(for $h=\pi^+,\ \pi^-$) 
for $\Delta u/u$, $\Delta d/d$ and $\Delta
  q_s/q_s$. In this analysis we treat the "data" as perfectly known,
  but include realistic errors on the purities, arising from the errors
  on our FFs. In this way we illustrate the uncertainty on the polarized
  parton densities arising solely from the uncertainties on the purity functions.

  The results are shown in Fig.~\ref{fig7}.
\figg 
It is seen that whereas $\Delta
  u/u$ is largely insensitive to the uncertainty on the purity,
  both $\Delta d/d$ and $\Delta q_s/q_s$ inherit significant errors from this
  uncertainty.

Bearing in mind that the errors shown in
Fig.~\ref{fig7} arise solely from the uncetainty on the purities, one learns from
this study that it is misleading to treat the purities as absolutely known
quantities.
It would
  be far more meaningful to follow the strategy suggested in
  \cite{we2} and use the SIDIS data to obtain both the FFs and the
  polarized parton densities. The purity is an unnecessary element
  and in any case loses its usefulness in NLO.


\section{Conclusions}

   We have shown that a judicious combination of the HERMES  SIDIS data on
   $\pi^\pm$ production
   and certain aspects of the data on $e^+e^- \to \pi^\pm X$ allows
   the extraction, for the first time,
   of the flavour separated fragmentation functions $D^{\pi^+}_u$, $D^{\pi^+}_d$ and
   $D^{\pi^+}_s$.

  The key element in this approach is the avoidance of {\it any ad
  hoc}
  model-dependent flavour separation in the $e^+e^-$ data,
  by noting that at the $Z^0$ peak what is well determined is essentially
  the light flavour singlet combination of fragmentation functions
  $D_\Sigma^{\pi^+}=2\left( D_u^{\pi^+}+  D_d^{\pi^+}+  D_s^{\pi^+}\right)$,
  which is almost identical in all analyses of the $e^+e^-$ data.
  The negative aspect of this approach is the need to evolve $D_\Sigma^{\pi^+}$ down
  from the $Z^0$ region to the SIDIS region of a few $(GeV)^2$,
  which involves mixing with the poorly known gluon fragmentation function.

  In fact the HERMES data is extremely accurate, so that almost all
  the uncertainty in our determination of $D_u^{\pi^+}$, $D_d^{\pi^+}$  and
  $D_s^{\pi^+}$
  arises from the uncertainty in the evolution of $D_\Sigma^{\pi^+}$.
  As it turns out, this has little effect on $D_u^{\pi^+}$
  and $D_d^{\pi^+}$, which are
  very well determined, but $D_s^{\pi^+}$ has relatively large errors.

   We have also examined the question of the precision with which
   the polarized parton densities could be extracted from future polarized
   SIDIS pion production data. Here we have assumed perfect 'data',
   then followed the HERMES purity method to obtain the polarized parton
   densities, and thereby displayed
the uncertainties generated solely
by the
   errors on the purity functions. The significance of this study
   is that in the earlier analyses \cite{SMC,HERMES} the purity functions are taken
   as almost perfectly known with essentially no errors. As expected we
   have found that  $\Delta d$ and $\Delta q_s$ are significantly affected by
   the uncertainties in the purity functions. This suggests that in the published
   polarized parton densities
 extracted from polarized SIDIS $h^\pm$-production data \cite{SMC,HERMES,deFlorian},
the uncertainties given are missing an inherent
error arising from the fragmentation uncertainties as quantified in this paper.


\section{Acknowledgements}
  We are grateful to Mark Beckmann and Helmut B\"ottcher for many stimulating
  discussions. This research was supported by a UK Royal Society Collaborative Grant
and by the
National Science Foundation under Grant PHY-0070443.

 \newpage

\end{document}